# Dynamics of electrochemical flows I: Motion of electrochemical flows

Chengjun Xu[1*], Chin-Tsau Hsu[1,2]

[1]Graduate School at Shenzhen, Tsinghua University, Shenzhen City, Guangdong Province, 518055, China. E-mail address: vivaxuchengjun@163.com.

[2]Department of Mechanical Engineering, Hong Kong University of Science and Technology, Kowloon, Hong Kong. E-mail address: mecthsu@ust.hk (CT).

**Abstract**: The motion of the electrolyte, comprising of solute ions and solvent molecules, is a frequently-occurring natural phenomenon. The motion of the electrolyte leads to the flows of ions and solvent molecules, known as electrochemical flows. In this study, we establish a general theory to describe the motion of the electrochemical flows. Our theory provides a different approach from others to clarify the details of the transport phenomena for the electrochemical flows. We derive the governing equations in the electrolyte fluid from mass, charge, momentum, energy, and concentration conservations. In addition, we normalize the governing equations to derive the dimensionless parameters, known as Reynolds, Thompson, Peclet, Prandtl and *X* numbers. The physical meaning of these parameter numbers in the electrochemical flow is discussed in detail. A new number, named *X* number, appears in the Navier-Stokes equation symbolizing the balance between the inertia force and the electric force.

1 Introduction

The motion of the electrolyte is a natural phenomenon, which frequently occurs. The electrolyte generally comprises of solute and solvent. The solute normally is dissolved in the solvent as cations and anions. The motion of the electrolyte leads to the flows of ions and solvent molecules, known as electrochemical flows. The streaming of the electrochemical flows occurs generally under a certain "driving force". For example, the solvent molecules can move through a permeable membrane into a region of higher solute concentration under an osmosis pressure. In electrophoresis the dispersed particles and ions move under the influence of a spatially uniformed electric field. In the flow battery, an electrolyte containing cations, anions and solvent molecules is pumped through an electrochemical cell that reversibly converts chemical energy directly into electricity. In addition, the motion of the electrochemical flows is the key step in the electrochemical systems, i.e. electrochemical reactors, batteries, supercapacitors, fuel cells, etc, in which the electrochemical flows have to flow through the solid phase to form the electrochemical double layer and/or to participate the electrochemical reactions [1-3].

The thermodynamics of the mixture electrolyte with arbitrary components was well established. In an electrolyte, the Coulomb force between the ions will affect the thermodynamic and other physical properties of the system; these effects can be calculated because the force is known. The mathematical difficulties involved were first overcome by Debye and Hückel [4]. They established the first mathematical model on the electrolyte by assuming an ion atmosphere and disregarding the solvent molecules. Individual ion is surrounded by a spherically symmetric cloud of other ions. The total (mean) electric charge of an ionic atmosphere (cloud) equals that of the central ion with opposite sign. In an undisturbed electrolyte, because the Columbic force effect and electrophoresis are proportional to the square root of the concentration, the physical properties, for example colligative properties, of the electrolyte are proportional to the square root of the concentration, on account of the factor $\kappa \sim \sqrt{C}$, where $\kappa^{-1}$ is Debye length. In the following L. Onsager considered the asymmetry of ion atmosphere and the force transfer from ions to solvents and derived a general limiting law to describe the physical properties of the electrolyte with arbitrary components [5]. For a disturbed electrolyte the motion of the individual central ion or ion atmosphere is generally described according to the Stoke's law. In the electrophoresis, von Smoluchowski developed a widely used theory to describe the mobility (the velocity divided by the applied field) of a dispersed particle in the fluid with a low Reynolds number [6-8].

The motion of electrochemical flow is an important process frequently encountered in nature or in engineering. However, the existing theories mainly focus on one part of the motion of the ions or solvent of the electrolyte[6-8]. It generally believes that if one species in the electrolyte start moving under the certain driving force, the other species will be affected due to the collision among the molecules and ions. The description of the electrochemical flows needs the general clarification of the details of the transport phenomena, for example the transference of the mass, heat and electric force and the velocity of the ions and solvents. In addition, the fundamental equations occurring in the natural and physical sciences have to be obtained from conservation laws. Conservation laws are just balance laws, with equations to express the balance of some quantities throughout a process. Mathematically, conservation laws usually translate into integral/differential equations, which are then regarded as the governing equations of the process. Therefore the fundamental equations governing the electrochemical flows have to be derived from the conservation laws.

In this work, we try to establish a general theory to describe the electrochemical flows from charge, mass, momentum, energy, and concentration conservation laws. Our theory provides a different approach from others to clarify the details of the transport phenomena for the electrochemical flows. We will establish the conservation equations in the electrolyte fluid by means of establishing microscopic representative elementary volume. The microscopic governing equations will be derived from the local conservation laws in the electrolyte fluid. We further normalize the governing equations to derive the dimensionless parameters, known as Reynolds, Thompson, Peclet, Prandtl and *X* numbers. The physical meaning of these parameter numbers in the electrochemical flow is discussed in detail. A new *X* number appears in the Navier-Stokes equation symbolizing the balance between the inertia force and the electric force. In the next paper, the process of the electrochemical flows-through porous electrode will be addressed.

2 Definition of representative elementary volume

For the general situation, it is very difficult to follow the individual molecules because of the enormous numbers, while it is easy to collect a large number of readings of property at a fixed point (e.g. velocity of the particles passing it) and to take the average of the sum. Hence the technique of statics or averaging is necessary. Spatial averaging is to average a quantity over a group of particles in a controlled volume. Statics uses this controlled volume to represent the sample. One controlled volume can contain a lot of molecules. In the controlled volume there are many molecules with different properties. But for the controlled volume its properties are averaged by the properties of all the molecules. Therefore, the controlled volume can be used to represent the sample and called as representative elementary volume (REV). The whole sample we study can be seen to be consisted of continuous REVs. If the REV is centered at a point $\mathbf{r}(x_1, x_2, x_3)$ at Cartesian coordinated system, the physical quantities (values) of material properties will change as REV excurses in space of different $\mathbf{r}$. Then a physical property *f* becomes a function of ($\mathbf{r}$,*t*), which is analytical, i.e. continuous, differentiable and integrable. In the whole domain (field), where the physical property is defined, the whole field becomes continuum. By using the REV, the Lagrangian of physical quantity (*f*) of the individual molecule is transferred into Eurlerian of a group of molecules. The relation between Lagrangian expression and Eulerian expression is written as:

$$\frac{df}{dt} = \frac{\partial f}{\partial t} + (\mathbf{v} \cdot \nabla) f \qquad (1)$$

where **v** is the velocity and ∇ is the Hamilton operator.

In order to describe the motion of ions and solvent molecules, we establish a microscopic REV in the electrolyte fluid. Figure 1 shows the schematic picture of microscopic REV in the domain of electrolyte fluid.

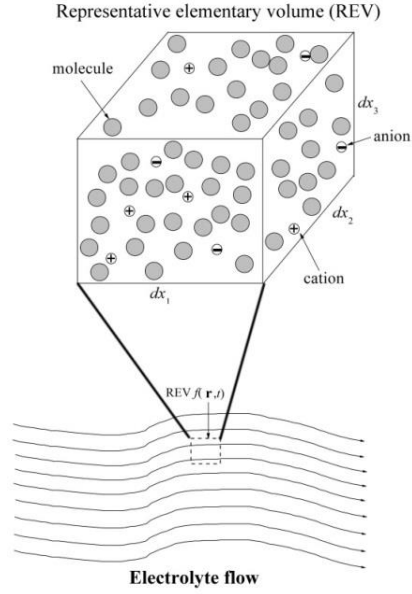

Figure 1 The schematic picture of a microscopic representative volume (REV)

3 Definition of physical quantity in microscopic REV

Considering one REV in the electrolyte fluid as shown in Figure 1, there are individual groups of solvents and ions. After we established the REV, we can define the physical quantities and investigate the conservation laws in the REV without ambiguous.

In the REV, we now consider the electrolyte composed of multi-component ions and solvent molecules, with mass $m_i$ where $i=1,2,3…,M$, with the molecular number $N_i$, i.e. $N = N_1 + N_2 + N_3 + ... + N_i + … + N_M$.

The number density in REV is

$$n = \frac{N}{V_c} = \frac{N_1}{V_c} + \frac{N_2}{V_c} + \frac{N_3}{V_c} + \cdots + \frac{N_i}{V_c} + \cdots + \frac{N_M}{V_c}$$

$$= n_1 + n_2 + n_3 + \cdots + n_i + \cdots + n_M \qquad (2)$$

It is clearly that the mass density for species *i* is

$$\rho_{mi} = \frac{1}{V_c}\sum_i^{N_i} m_i = n_i m_i = C_i M_i \quad (3)$$

The total mass density is

$$\rho_m = \rho_{m1} + \rho_{m2} + \rho_{m3} + \cdots \rho_{mi} + \cdots + \rho_{mM} = \sum_i^M \rho_{mi} \quad (4)$$

The charge density of species $i$ is

$$\rho_{ci} = \frac{1}{V_c}\sum_i^{N_i} z_i q_e = z_i F \rho_{mi}/M_i = z_i F C_i \quad (5)$$

where $q_e$ is the elementary charge, $M_i$ is the atomic weight of species $i$, $C_i$ is the concentration of species $i$, $F$ is the Faradic constant and $z_i$ is the charge number of the ion. For the solvent, the charge number is zero.

The total charge density is

$$\rho_c = \sum_i^M \rho_{ci} \quad (6)$$

The momentum for species $i$ is

$$\mathbf{M}_i = \rho_{mi}\mathbf{u}_i \quad (7)$$

The average velocity of species $i$ is

$$\mathbf{u}_i = \mathbf{M}_i/\rho_{mi} \quad (8)$$

The total momentum is

$$\mathbf{M} = \rho_m \mathbf{u} = \sum_i^M \rho_{mi}\mathbf{u}_i \quad (9)$$

The total average velocity is

$$\mathbf{u} = \mathbf{M}/\rho_m \quad (10)$$

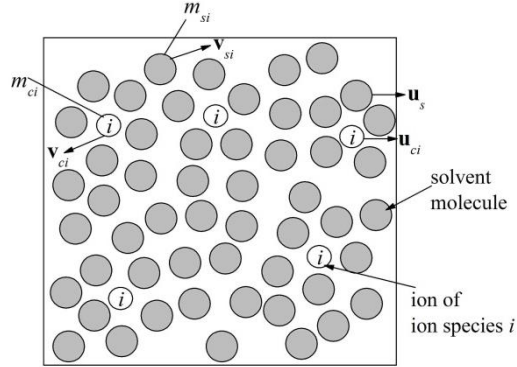

Figure 2 Schematic image of information of ion and solvent in the representative elementary volume (REV). For simplification, we only draw one kind of solvent molecules denoted as subscript "*s*" and one kind of ion species *i* denoted as subscript "*ci*".

After the definition of physical quantities, now we consider the interaction among species to derive the temperature and stress. One facet of the REV presenting in the electrolyte fluid is shown in Figure 2. There are individual electrochemical flows of solvents and ions. Ions and molecules normally collide from each other in the movement. Normally, the electrolyte can be seen as the dissolution of solute in the solvent. Once the solute dissolved in the solvent, the ions of solute will be surrounded by the solvent molecules to form solvation of the ions. Therefore, the ions are assumed to be distributed in the solvent matrix. We also further assume that the solvent is the matrix in the normal electrolyte and the relative permittivity of the electrolyte is constant. For a highly concentrated electrolyte with arbitrary components, in which the solvent cannot be seen as the matrix, we will discuss it in a latter paper. If one species moves under an external drag force, the other species will start moving by the collision among species.

The size of the REV is much larger than the size of the ion or molecule. Therefore, the solvation and the ion atmosphere of the ions are considered. The ion locates in the central surrounding by the solvent molecules and other ions. According to the model of P. Debye and L. Onsager, the presence of the electric force between ions will affect the chemical potential of the ion, which influences the velocity of the ion. In the following, we will define the temperature and stress tensor for the mixture of the electrolyte according to the collision among ions and molecules with different velocities.

The thermal velocities of each component are defined as

$$\mathbf{v}'_{i\alpha} = \mathbf{v}_{i\alpha} - \mathbf{u}_i \quad (11)$$

then

$$\sum_{\alpha}^{N_i} \mathbf{v}'_{i\alpha} = 0 \quad (12)$$

where the subscript $\alpha$ symbolizes one particle of species $i$ and $\mathbf{v}_{i\alpha}$ is the velocity of one particle of species $i$.

The total stress tensor is

$$\boldsymbol{\sigma} == \frac{1}{V_c}\sum^{N} m_i \mathbf{v}'_{i\alpha} \mathbf{v}'_{i\alpha} = \sum_{i}^{M}[\boldsymbol{\sigma}_i^{\ominus} + \rho_{mi}(\mathbf{u}_i - \mathbf{u})(\mathbf{u}_i - \mathbf{u})] \quad (13a)$$

where $\boldsymbol{\sigma}_i^{\ominus}$ is the standard stress tensor of each component.

$$\boldsymbol{\sigma}_i^{\ominus} = \frac{\rho_{mi}}{N_i}\sum_{\alpha}^{N_i} \mathbf{v}'_{i\alpha} \mathbf{v}'_{i\alpha} \quad (13b)$$

The stress tensor for species $i$ is

$$\boldsymbol{\sigma}_i = \boldsymbol{\sigma}_i^{\ominus} + \rho_{mi}(\mathbf{u}_i - \mathbf{u})(\mathbf{u}_i - \mathbf{u}) \quad (14)$$

The pressure for the ideal electrolyte mixture is

$$p = \frac{1}{3}tri(\boldsymbol{\sigma}) = CRT\dagger \quad (15)$$

where $C$ is the total molar concentration, $R$ is the Avogadro constant, and † is Van't Hoff factor.

The partial pressure for species $i$ is

$$p_i = x_i p \quad (16)$$

where $x_i$ is the molar fraction of species $i$.

The total kinetic energy per unit volume is

$$\frac{1}{V_c}\sum_{i}^{N}\frac{1}{2}m_i \mathbf{v}_i \cdot \mathbf{v}_i = \frac{3}{2}nkT + \frac{1}{2}\rho_m \mathbf{u} \cdot \mathbf{u} = \sum_{i}^{M}\frac{1}{2}\rho_{mi}\mathbf{u}_i \cdot \mathbf{u}_i + \sum_{i}^{M}\frac{3}{2}n_i kT_i \quad (17)$$

Here the internal kinetic energies per unit volume for each component are:

$$\frac{3}{2}nkT_i = \frac{1}{V_c}\sum_{\alpha}^{N_i}\frac{1}{2}m_i \mathbf{v}'_{i\alpha} \cdot \mathbf{v}'_{i\alpha} \quad (18)$$

where $T_i$ and $T$ are the temperatures of individual component $i$ and the total mixture, respectively. Therefore,

$$\frac{3}{2}nkT = \sum_{i}^{M}\frac{1}{2}\rho_{mi}\mathbf{u}_i \cdot \mathbf{u}_i + \sum_{i}^{M}\frac{3}{2}n_i kT_i - \frac{1}{2}\rho_m \mathbf{u} \cdot \mathbf{u} \quad (19)$$

This equation defines the temperature of the mixture. Alternatively we can derive the temperature directly as:

$$\frac{3}{2}nkT = \sum \frac{1}{V_c}\sum_{\alpha}^{N_i}\frac{1}{2}m_i (\mathbf{v}_{i\alpha} - \mathbf{u}) \cdot (\mathbf{v}_{i\alpha} - \mathbf{u})$$

$$= \sum_{i}^{M}\frac{3}{2}n_i kT_i + \sum_{i}^{M}\frac{1}{2}\rho_{mi}(\mathbf{u}_i - \mathbf{u}) \cdot (\mathbf{u}_i - \mathbf{u})$$

$$= \sum_{i}^{M}\frac{1}{2}\rho_{mi}\mathbf{u}_i \cdot \mathbf{u}_i + \sum_{i}^{M}\frac{3}{2}n_i kT_i - \frac{1}{2}\rho_m \mathbf{u} \cdot \mathbf{u} \quad (20)$$

When at thermal equilibrium, we get

$$T = T_1 = T_2 = T_3 = \cdots = T_i = \cdots = T_M \tag{21}$$

$$\mathbf{u} = \mathbf{u}_1 = \mathbf{u}_2 = \mathbf{u}_3 = \cdots = \mathbf{u}_i = \cdots = \mathbf{u}_M \tag{22}$$

The term $\rho_{mi}(\mathbf{u}_i\text{-}\mathbf{u})\cdot(\mathbf{u}_i\text{-}\mathbf{u})/2$ is the macroscopic kinetic energy per unit volume of the species *i* due to the deviation from the velocity **u** at equilibrium. Hence, it is considered as part of the internal kinetic energy of the mixture system since it will eventually be converted into internal thermal energy of the mixture when reaching equilibrium.

At the thermal equilibrium, the mean group velocities of the ions and the solvents are equal. Note that we use the REV to build the continuity of the interested domain. Such treatment is interested in the group quantities and REV contains lots of ions and solvent molecules. Therefore, the solvation of ion can be seen as the ion in the center having a mean velocity with the solvent molecules around. The fluctuation velocity of ion and solvent molecules will be transferred into the heat by collision in the REV. This assumption guarantees the reasonable explanation of the solvation and de-solvation in the electrochemical systems, for example in the supercapacitors [9]. In addition, the ion atmosphere of each ion is also included in the REV since the REV contains hundreds of thousands of ions. Because the mean velocities for each ion species are equal, the ion atmosphere for every ion in the REV can be seen as symmetry during movement. The detailed discussion will be addressed in a latter paper.

4 Governing equations in microscopic REV

The fundamental equations occurring in the natural and physical sciences are obtained from conservation laws. Conservation laws are just balance laws, with equations to express the balance of some quantities throughout a process. Mathematically, conservation laws usually translate into integral/differential equations, which are then regarded as the governing equations of the process. Therefore, we shall formulate the conservation laws into their governing equations mathematically. There are four major conservation laws in continuum transport phenomena. There are mass conservation, momentum conservation, energy conservation, and mass concentration conservation. For the electrochemical flows, the charge is also conserved.

After the carefully definition of physical quantities by the concept of controlled volume, let's move back to the electrochemical flows. In order to get the governing equations, any physical quantity (*f*) of the electrochemical flows is written as *f*(*x*₁, *x*₂, *x*₃,*t*) at a REV **r**(*x*₁, *x*₂, *x*₃). The governing equations will get from the conservation of charge, mass, momentum, energy and concentration in the REV.

4.1 Conservation of mass

According to the definition of the density, the total mass in the controlled volume REV, which encloses fluid particles in motion, is given by:

$$m = \int_V \rho_m dV \qquad (23)$$

The local mass conservation in the REV requires $dm/dt = \dot{m}$, and

$$\dot{m} = \frac{dm}{dt} = \frac{d}{dt}\left[\int_V \rho_m dV\right] = \int_V \frac{\partial \rho_m}{\partial t} dV + \int_S \rho_m \mathbf{u} \cdot d\mathbf{s} \qquad (24)$$

where $\dot{m}$ is the source of the mass. This is integral form of the conservation of mass. The last term on the right hand side of above equation is the rate of mass change due to the change of $V_c$ by the deformation of surface *S* that enclosed $V_c$. If there is no the mass source in the REV, by applying the divergence theorem, the conservation of mass is rewritten as

$$\frac{\partial \rho_m}{\partial t} + \nabla \cdot (\rho_m \mathbf{u}) = 0 \qquad (25)$$

The electrolyte can be seen as incompressible fluid, which means that $\rho_m$ is constant. Hence the mass conservation equation becomes:

$$\nabla \cdot \mathbf{u} = 0 \qquad (26)$$

For the species *i*, the group cannot be seen as incompressible fluid. Therefore, we get

$$\frac{\partial \rho_{mi}}{\partial t} + \nabla \cdot (\rho_{mi}\mathbf{u}_i) = \dot{m}_{ci} \qquad (27)$$

where $\dot{m}_{ci}$ is the mass source of species *i* in the REV. If there is no mass source this term is zero. By using $C_i$, we get

$$\frac{\partial C_i}{\partial t} + \nabla \cdot (C_i \mathbf{u}_i) = \dot{m}_{ci} \quad (28)$$

For the solvent, we assume that the solvent is the matrix, which means the solvent is incompressible. Therefore, we get

$$\nabla \cdot \mathbf{u}_s = 0 \quad (29)$$

where the subscript "$s$" represents the solvent.

4.2 Charge conservation

According to the definition of charge density, the total charge ($Q_i$) of species $i$ in the REV is

$$Q_i = \int_V \rho_{ci} dV \quad (30)$$

For the species $i$, the charge conservation can be obtained from the Guass's law in the REV:

$$\int_S \mathbf{E}_{ri} \cdot d\mathbf{s} = \int_V \frac{\rho_{ci}}{\varepsilon} dV \quad (31)$$

where $\mathbf{E}_{ri}$ is the electric flied generated by the species $i$ and $\varepsilon$ is the dielectric constant of the electrolyte. For the solvent, the charge density is equal to zero.

And since this holds for any REV, the differential form of Guss's law writes

$$\nabla \cdot \mathbf{E}_{ri} = \frac{\rho_{ci}}{\varepsilon} = \frac{z_i F C_i}{\varepsilon} = -\nabla^2 \Phi_{ri} \quad (32)$$

where $\Phi_{ri}$ is the potential of ion species $i$.

The total electric flied ($\mathbf{E}_r$) generated in the REV is

$$\mathbf{E}_r = \sum_i^n \mathbf{E}_{ri} \quad (33)$$

By applying the superposition law the electric flied in the REV is

$$\nabla \cdot \mathbf{E}_r = \frac{\rho_c}{\varepsilon} = -\nabla^2 \Phi_r = 0 \qquad (34)$$

where $\mathbf{E}_r$ and $\Phi_r$ are the electric flied generated inside the REV and the potential of the REV, respectively. Normally in the electrolyte the total charge density is equal to zero due to the neutrality.

4.3 Conservation of momentum

The Newton's second law is stated as:

$$\mathbf{F} = \frac{d\mathbf{M}}{dt} = \frac{d}{dt}\left[\int_V \rho_m \mathbf{u} dV\right] = \int_V \frac{\partial}{\partial t}(\rho_m \mathbf{u})dV + \int_S \rho_m \mathbf{u}\mathbf{u} \cdot d\mathbf{s} \qquad (35)$$

Hence the neat force includes the surface force due to the surface stresses (σ) and the body force (b) due to the flied force (i.e. electric, gravity forces, etc.). With the vicious stress, the total stress on the fluid particle (REV) is the sum of pressure stress ($\sigma_p = -p\mathbf{I}$, here the negative sign implies that tension is positive) and the vicious stress (τ).

$$\boldsymbol{\sigma} = \boldsymbol{\sigma}_p + \boldsymbol{\tau} = -p\mathbf{I} + \mu \mathbf{S} + (\varsigma - \frac{2}{3}\mu)(\nabla \cdot \mathbf{u})\mathbf{I} \qquad (36)$$

where $\mu$ is the dynamic shear viscosity of the electrolyte, $\mathbf{S}$ is the strain rate tensor, ς is the volume viscosity, and $\mathbf{I}$ is the unit tensor.

$$\mathbf{I} = \begin{bmatrix} 1 & 0 & 0 \\ 0 & 1 & 0 \\ 0 & 0 & 1 \end{bmatrix} \qquad (37)$$

The strain rate tensor $\mathbf{S}$ is a symmetric tensor that measures the rate of linear and angular deformations of fluid element. The strain rate tensor is expressed as:

$$S = \nabla v + (\nabla v)^T = \begin{bmatrix} 2\partial u/\partial x & (\frac{\partial u}{\partial y}+\frac{\partial v}{\partial x}) & (\frac{\partial u}{\partial z}+\frac{\partial w}{\partial x}) \\ (\frac{\partial u}{\partial y}+\frac{\partial v}{\partial x}) & 2\partial v/\partial y & (\frac{\partial v}{\partial z}+\frac{\partial w}{\partial y}) \\ (\frac{\partial u}{\partial z}+\frac{\partial w}{\partial x}) & (\frac{\partial v}{\partial z}+\frac{\partial w}{\partial y}) & 2\partial w/\partial z \end{bmatrix} \quad (38)$$

where $\mathbf{v}(u,v,w)$ is the velocity vector and the superscript "$T$" represents the transpose.

The total neat force on the REV is given by

$$\mathbf{F} = \int_V \frac{\partial}{\partial t}(\rho_m \mathbf{u})dV + \int_S \rho_m \mathbf{uu} \cdot dS = \int_S \boldsymbol{\sigma} \cdot ds + \int_V \mathbf{b}dV \quad (39)$$

The differential form of the momentum conservation is written as

$$\frac{\partial}{\partial t}(\rho_m \mathbf{u}) + \nabla \cdot (\rho_m \mathbf{uu}) = -\nabla p + \nabla \boldsymbol{\tau} + \mathbf{b} \quad (40)$$

In an electrochemical system, the liquid electrolyte can be seen as incompressible fluid ($\nabla \mathbf{u} = 0$).

Therefore we get

$$\rho_m \left[ \frac{\partial \mathbf{u}}{\partial t} + \nabla \cdot (\mathbf{uu}) \right] = -\nabla p + \mu \nabla^2 \mathbf{u} + \rho_c \mathbf{E} + \rho_m \mathbf{g} \quad (41)$$

For the species $i$, we get

$$\frac{\partial}{\partial t}(\rho_{mi} \mathbf{u}_i) + \nabla \cdot (\rho_{mi} \mathbf{u}_i \mathbf{u}_i) = \nabla \boldsymbol{\sigma}_i + \rho_{ci} \mathbf{E} + \rho_{mi} \mathbf{g} \quad (42)$$

where **g** is gravity acceleration. It has to point that the electric flied **E** is the total electric flied. We ignore the gravity for species. We further assume that the group of species $i$ is moving at its maximum velocity (known as terminal velocity),

$$-\nabla p_i + \mu \nabla^2 \mathbf{u}_{ci} + z_i F C_i \mathbf{E} = 0 \quad (43)$$

This equation indicates that the electric force is balanced with the viscous force. For the solvent, because it is electroneutral and we assume it is the matrix we get

$$\rho_{ms}[\frac{\partial}{\partial t}\mathbf{u}_s + \nabla \cdot (\mathbf{u}_s\mathbf{u}_s)] = -\nabla p_s + \mu\nabla^2\mathbf{u}_s + \rho_{ms}\mathbf{g} \qquad (44)$$

where the subscript "*s*" represents the solvent. Note that the solvent is assumed to be the matrix and the density of the solvent is assumed to be constant.

4.4 Conservation of energy

At the thermal equilibrium, the temperature of each group in the REV is constant. For the integral electrolyte, we get

$$\int_V \frac{\partial}{\partial t}(\rho_m C_p T)dV + \int_S \rho_m C_p T\mathbf{u} \cdot dS = -\int_S k\nabla T \cdot dS + \int_V H_r dV \qquad (45)$$

It is in the differential form

$$\rho_m C_p[\frac{\partial T}{\partial t} + \nabla \cdot (T\mathbf{u})] = -\nabla \cdot (k\nabla T) + H_r \qquad (46)$$

where $C_p$ and $k$ are the thermal capacity and the thermal conductivity of the electrolyte fluid. $H_r$ is the heat generated inside the REV.

For the species *i*, at thermal equilibrium we get

$$T = T_i \qquad (47)$$

Therefore, it is no necessary to build an equation for the each species.

4.5 Conservation of mass concentration

For the species *i*, the molar concentration ($C_i$) conservation is considered as

$$\frac{\partial}{\partial t}(\rho_m C_i) + \nabla \cdot (\rho_m C_i \mathbf{u}_i) = -\nabla \cdot N_i + \dot{m}_{ci} \qquad (48)$$

where $N_i$ is the mass diffusion flux. The $N_i$ in the electrochemical system is:

$$N_i = -\lambda_i \nabla C_i + \frac{z_i F}{RT} \lambda_i C_i \mathbf{E} \quad (49)$$

where $\lambda_i$ is the diffusion conductivity and $R$ is the Avogadro constant. Hence the concentration conservation is rewritten as

$$\rho_m \left[\frac{\partial C_i}{\partial t} + \nabla \cdot (C_i \mathbf{u}_i)\right] = \nabla \cdot \left(\lambda_i \nabla C_i - \frac{z_i F \lambda_i}{RT} C_i \mathbf{E}\right) + \dot{m}_{ci} \quad (50)$$

Note that there is a relation between $D_i$ and $\lambda_i$.

$$D_i = \frac{\lambda_i}{\rho_m} \quad (51)$$

By assuming the diffusion coefficient is isotropic, the concentration conservation for species $i$ can be rewritten as

$$\frac{\partial C_i}{\partial t} = D_i \nabla^2 C_i - \frac{z_i F D_i}{RT} \nabla \cdot (C_i \mathbf{E}) - \nabla \cdot (C_i \mathbf{u}_i) + \dot{m}_{ci} \quad (52)$$

4.6 Current equation

Considering the continuity of REV in the flied, the motion of species $i$ with a mean velocity ($\mathbf{u}_i$) gives the total current flux as:

$$\mathbf{J}_i = -z_i F D_i \nabla C_i + \frac{z_i^2 F^2}{RT} D_i C_i \mathbf{E} + z_i F C_i \mathbf{u}_i \quad (53)$$

The total current $\mathbf{J}$ is

$$\mathbf{J} = \sum_i^n \left[-z_i F D_i \nabla C_i + \frac{z_i^2 F^2}{RT} D_i C_i \mathbf{E} + z_i F C_i \mathbf{u}_i\right] \quad (54)$$

where $D_i$ is the mass diffusion coefficient and $\mathbf{E}$ is the total electric flied through the REV surface. The summary of the governing equations for the integral electrolyte is listed in the Table 1. Table 2 summarizes the governing equations for species $i$.

Table 1 Summary of the governing equations for the integral electrolyte

| Name | Equation |
|---|---|
| Charge conservation | $-\nabla^2 \mathbf{\Phi}_r = \dfrac{\rho_c}{\varepsilon}$ |
| Mass conservation | $\nabla \cdot \mathbf{u} = 0$ |
| Momentum conservation | $\rho_m \dfrac{\partial \mathbf{u}}{\partial t} + \rho_m \nabla \cdot (\mathbf{u}\mathbf{u}) = -\nabla p + \mu \nabla^2 \mathbf{u} + \rho_c \mathbf{E} + \rho_m \mathbf{g}$ |
| Energy conservation | $\rho_m C_p \dfrac{\partial T}{\partial t} + \rho_m C_p \nabla \cdot (T\mathbf{v}) = -\nabla \cdot (k\nabla T) + \mathrm{H}_r$ |

Table 2 Summary of the governing equations for the species $i$

| Name | Equation |
|---|---|
| Charge conservation | $\nabla \cdot \mathbf{E}_{ri} = \dfrac{z_i F C_i}{\varepsilon}$ |
| Mass conservation | $\dfrac{\partial C_i}{\partial t} + \nabla \cdot (C_i \mathbf{u}_i) = \dot{m}_{ci}$ |
| Momentum conservation | $\dfrac{\partial}{\partial t}(\rho_{mi} \mathbf{u}_i) + \nabla \cdot (\rho_{mi} \mathbf{u}_i \mathbf{u}_i) = \nabla \boldsymbol{\sigma}_i + \rho_{ci} \mathbf{E} + \rho_{mi} \mathbf{g}$ |
| Concentration conservation | $\rho_m \left[\dfrac{\partial C_i}{\partial t} + \nabla \cdot (C_i \mathbf{u}_i)\right] = \nabla \cdot \left(\lambda_i \nabla C_i - \dfrac{z_i F \lambda_i}{RT} C_i \mathbf{E}\right) + \dot{m}_{ci}$ |
| Current equation | $\mathbf{J}_i = -z_i F D_i \nabla C_i + \dfrac{z_i^2 F^2}{RT} D_i C_i \mathbf{E} + z_i F C_i \mathbf{u}_i$ |

5 Normalization of governing equations

Normalization is a method to analyze governing equations for certain physical phenomenon, by substituting properly scaled variables. This procedure removes the dependence on units in the equations. Normalization also makes the variables in the equations to become order one, i.e. O(1), to facilitate comparisons of relative importance of each terms in the equations. Characteristics of the equation at certain scales (chosen by the observer) emerge in form of non-dimensional parameters, often relating to length, time, and force quantities. Flows are said to be dynamically similar if scaling the dependent and independent variables yield the same non-dimensional parameter in the governing equations.

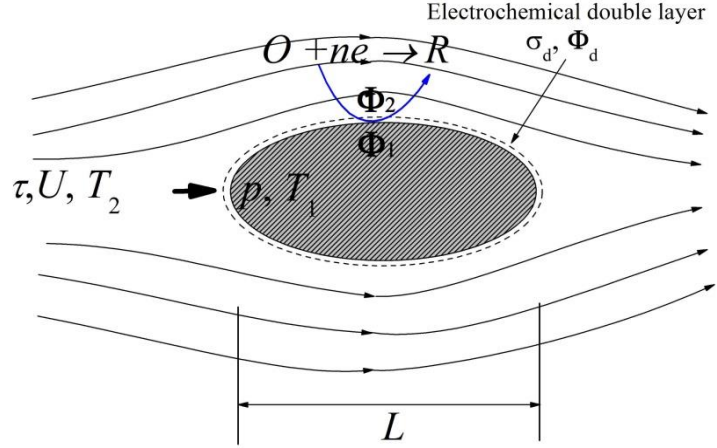

Figure 3 Electrolyte flows through one solid domain

In the electrolyte fluid the electrochemical flows pass through the solid phase to enable occurrence of the electrochemical reaction and the electrochemical double layer. Figure 3 shows the schematic picture that the electrochemical flows pass through solid phase. One electrochemical fluid lasting in a time scale of $\tau$ flows through one solid particle at a landscape of $L$. The original velocity and temperature of the electrochemical fluid is $U$ and $T_2$. When it passes through the solid phase, the fluid will trigger the pressure ($P$) on the surface of the solid phase toward it. The electrochemical double layer and reaction occur at the interface between the solid and the electrochemical fluid during the time scale of $\tau$. $\Phi_1$ and $\Phi_2$ are potential of solid phase and fluid, respectively. $\Phi_d$ and $\sigma_d$ are the potential and surface charge density of the electrochemical double layer at the interface between solid and fluid. $\eta$ is the overpotential related to the electrochemical reaction. The subscript 1 and 2 mean the solid and fluid phase, respectively. Before the normalization, we have to select the proper scales. Considering the flow-through the solid phase as shown in Figure 3, the proper scales are $\tau$ for time, $U$ for velocity, $P$ for pressure, $L$ for the length, $T_1$ and $T_2$ for temperature, $\Phi_\Theta$ for potential, $C_i^0$ for concentration, and $\mathbf{J}_0$ for current. $T_1$ is the temperature of solid phase. $\mathbf{J}_0$ is the exchange current flux at the interface due to the electrochemical reaction. $\Phi_\Theta$ is the reference potential, which you choose. Changing reference potential only amounts to adding a constant $K$ to the potential. $C_i^0$ is the concentration of species $i$ in the bulk electrolyte.

Table 3 summarizes the proper scales for the electrochemical flows. In the flow domain it is assumed that the length scales in $x_1$, $x_2$, and $x_3$ directions are the same. Using these scales, the variables are normalized to obtain the non-dimensional variables. For example, the new variable of time ($t^*$) is defined by the original one ($t$):

$$t^* = t/\tau \qquad (55)$$

where the superscript "*" represents the new variable.

Table 3 Summary of the proper scales and variables

| Name | Scale | Original variable | New variable |
|---|---|---|---|
| Time | $\tau$ | $t$ | $t^* = t/\tau$ |
| Velocity | $U$ | $\mathbf{u}$ | $\mathbf{u}^* = \mathbf{u}/U$ |
| Pressure | $P$ | $p$ | $p^* = p/P$ |
| Length | $L$ | - | - |
| Temperature | $T_1, T_2$ | $T$ | $T^* = \dfrac{T - T_2}{T_1 - T_2}$ |
| Potential | $\Phi_\Theta$ | $\Phi, \Phi_1, \Phi_2, \eta$ | $\Phi^* = \dfrac{\Phi}{\Phi_\Theta}$ |
| Current | $\mathbf{J}$ | $\mathbf{J}$ | $\mathbf{J}^* = \mathbf{J}/\mathbf{J_0}$ |
| Concentration | $C_i$ | $C_i$ | $C^* = C_i/C_i^0$ |

5.1 Normalization of mass conservation

After definition of dimensionless variables, the normalized governing equations will be derived. The mass conservation of the electrolyte becomes:

$$\frac{U}{L}\nabla^* \cdot \mathbf{u}^* = 0 \qquad (56)$$

Therefore,

$$\nabla^* \cdot \mathbf{u}^* = 0 \qquad (57)$$

Mass conservation of species $i$ is

$$\frac{C_i^0}{\tau}\frac{\partial C_i^*}{\partial t} + \frac{U}{L}C_i^0 \nabla^* \cdot (C_i^* \mathbf{u}_i^*) = \dot{m}_{ci} \qquad (58)$$

We can define the advective time ($\tau_a$), which indicates the time of flow-through the landscape of $L$.

$$\tau_a = \frac{L}{U} \qquad (59)$$

*Th* is the Thompson number.

$$Th = \frac{\tau_a}{\tau} \qquad (60)$$

Mass conservation of species *i* becomes

$$\frac{\partial C_i^*}{\partial t} + \frac{1}{Th}\nabla^* \cdot (C_i^* \mathbf{u}_i^*) = \dot{m}_{ci} \qquad (61)$$

5.2 Normalization of charge conservation

The normalization of charge conservation for the species *i* is

$$\nabla^{*2}\mathbf{\Phi}_{ri}^* = -\frac{L^2 z_i F C_i^0}{\varepsilon \Phi_\Theta} C_i^* \qquad (62)$$

For the electrolyte, the normalization of charge conservation is

$$\nabla^{*2}\mathbf{\Phi}_r^* = -\sum_i^M \frac{L^2 C_i^0 z_i F C_i^*}{\Phi_\Theta \varepsilon} \qquad (63)$$

5.3 Normalization of current equation

The current equation becomes

$$\mathbf{J}_0 \mathbf{J}_i^* = -\frac{z_i F C_i^0 D_i}{L}\nabla^* C_i^* - \frac{z_i^2 F^2 D_i}{RT}\frac{C_i^0 \Phi_\Theta}{L} C_i^* \nabla^* \mathbf{\Phi}_2^* + z_i F C_i^0 U C_i^* \mathbf{u}_i^* \qquad (64)$$

Both sides are divided by $z_i F C_i^0 U$.

$$\frac{\mathbf{J_0}}{z_i F C_i^0 U} \mathbf{J}_i^* = -\frac{D_i}{LU} \nabla^* C_i^* - \frac{D_i}{LU} \frac{z_i F \Phi_\Theta}{RT} C_i^* \nabla^* \mathbf{\Phi}_2^* + C_i^* \mathbf{u}_i^* \qquad (65)$$

5.4 Normalization of momentum conservation

The normalization of momentum conservation for integral electrolyte is

$$\rho_m \frac{U}{\tau} \frac{\partial \mathbf{u}^*}{\partial t^*} + \frac{U^2}{L} \rho_m \nabla^* \cdot (\mathbf{u}^* \mathbf{u}^*) = -\frac{P}{L} \nabla^* p^* + \frac{\mu U}{L^2} \nabla^{*2} \mathbf{u}^* + \rho_m \mathbf{g} \qquad (66)$$

We ignore the electric force term because the total charge density is normally equal to zero due to the electro-neutralization. The both sides are divided by $\rho_m U^2/L$:

$$\frac{1}{Th} \frac{\partial \mathbf{u}^*}{\partial t^*} + \nabla^* \cdot (\mathbf{u}^* \mathbf{u}^*) = -\frac{P}{\rho_m U^2} \nabla^* p^* + \frac{1}{Re} \nabla^{*2} \mathbf{u}^* + \frac{L}{U^2} \mathbf{g} \qquad (67)$$

where $Re$ is Reynolds number.

$$Re = \frac{\rho_m U L}{\mu} = \frac{UL}{\nu} \qquad (68)$$

where $\mu$ is the dynamic viscosity (kg m$^{-1}$ s$^{-1}$) and $\nu$ is the kinematic viscosity (m$^2$ s$^{-1}$)

For the solvent, the normalization of momentum conservation is

$$\frac{1}{Th} \frac{\partial \mathbf{u}_s^*}{\partial t^*} + \nabla^* \cdot (\mathbf{u}_s^* \mathbf{u}_s^*) = -\frac{P}{\rho_m U^2} \nabla^* p_s^* + \frac{1}{Re} \nabla^{*2} \mathbf{u}_s^* + \frac{L}{U^2} \mathbf{g} \qquad (69)$$

The momentum conservation for ion species $i$ is

$$-\nabla p_i + \mu \nabla^2 \mathbf{u}_i - z_i F C_i \mathbf{\Phi}_2 = 0 \qquad (70)$$

The normalization of momentum conservation for ion species $i$ is

$$-\frac{P}{\rho_m U^2} \nabla^* p_i^* + \frac{1}{Re} \nabla^{*2} \mathbf{u}_i^* - \frac{1}{X_i} C_i^* \nabla^* \mathbf{\Phi}_2^* = 0 \qquad (71)$$

$X_i$ number is a new number appearing in the Navior-Stokes equation

$$X_i = \frac{\rho_m U^2/L}{z_i F C_i^0 \Phi_\Theta /L} \tag{72}$$

5.5 Normalization of energy conservation

The energy equation only for the integral electrolyte becomes

$$\frac{1}{Th}\frac{\partial T^*}{\partial t^*} + \nabla^* \cdot (T^* \mathbf{u}^*) = \frac{1}{Pe}\nabla^2 T^* + H_r \tag{73}$$

where $Pe$ is Peclet number.

$$Pe = \frac{\rho_m C_p UL}{k} = \frac{UL}{\alpha} \tag{74}$$

$\alpha$ is the thermal diffusivity.

$$\alpha = \frac{k}{\rho_m C_p} \tag{75}$$

Note that the thermal conductivity is assumed to be isotropic in every direction.

5.6 Normalization of concentration conservation

The concentration equation becomes

$$\rho_m \left[\frac{C_i^0}{\tau}\frac{\partial C_i^*}{\partial t^*} + \frac{C_i^0 U}{L}\nabla^* \cdot (C_i^* \mathbf{u}_{ci}^*)\right] = \frac{\lambda_i C_i^0}{L^2}\nabla^{*2} C_i^* + \frac{z_i F \lambda_i}{RT}\frac{C_i^0 \Phi_\Theta}{L^2}\nabla^*(C_i^* \nabla^* \Phi_2^*) \tag{76}$$

Both sides are divided by $\rho_m C_0 U/L$.

$$\frac{1}{Th}\frac{\partial C_i^*}{\partial t^*} + \nabla^* \cdot (C_i^* \mathbf{u}_i^*) = \frac{1}{Pe_m}\nabla^{*2} C_i^* + \frac{z_i F \Phi_\Theta}{Pe_m RT}\nabla^*(C_i^* \nabla^* \Phi_2^*) \tag{77}$$

$Pe_m$ is the Peclet number for the mass transfer, which is the ratio between convective transfer and diffusive transfer.

$$Pe_m = \frac{\rho_m UL}{\lambda_i} = \frac{UL}{D_i} \tag{78}$$

# 6 Non-dimensional parameters

There are non-dimensional parameters appearing in the governing equations. These non-dimensional parameters carry the proper physical meanings. In the following, we will discuss the physical meaning of the parameters that we will use in the electrochemical flow.

## 6.1 Thompson number

Thompson number ($Th$) can be considered as the relative magnitude between two time scales, the advective time ($\tau_a$) and the response time ($\tau$).

$$Th = \frac{\tau_a}{\tau} \qquad (79)$$

$\tau_a$ is the advective time scale which measures the time required for the fluid particles to pass the body of size $L$.

When $Th \ll 1$, the problem is an unsteady or a transient problem. The system transient response time is shorter than the advective time scale, causing the transient term to be important.

On the other hand when $Th \gg 1$, the problem becomes essentially a steady problem (or sometime called as quasi-steady) at the considered time scale.

Mass conservation becomes:

$$\frac{\partial C_i^*}{\partial t} = \dot{m}_{ci} \qquad (80)$$

The equation of momentum conservation for the electrolyte becomes

$$\nabla^* \cdot (\mathbf{u}^* \mathbf{u}^*) = -\frac{P}{\rho_m U^2} \nabla^* p^* + \frac{1}{Re} \nabla^{*2} \mathbf{u}^* + \frac{L}{U^2} \mathbf{g} \qquad (81)$$

The equation of energy conservation becomes

$$\nabla^* \cdot (T^* \mathbf{u}^*) = \frac{1}{Pe} \nabla^{*2} T^* + H_r \qquad (82)$$

The equation of concentration conservation becomes

$$\nabla^* \cdot (C_i^* \mathbf{u}_i^*) = \frac{1}{Pe_m} \nabla^{*2} C_i^* + \frac{z_i F \Phi_\Theta}{Pe_m RT} \nabla^* (C_i^* \nabla^* \mathbf{\Phi}_2^*) \qquad (83)$$

## 6.2 Reynolds number

The Reynolds number ($Re$) is

$$Re = \frac{\rho_m UL}{\mu} = \frac{UL}{\nu} \qquad (84)$$

where $\mu$ is the dynamic viscosity (kg m$^{-1}$ s$^{-1}$) and $\nu$ is the kinematic viscosity (m$^2$ s$^{-1}$).

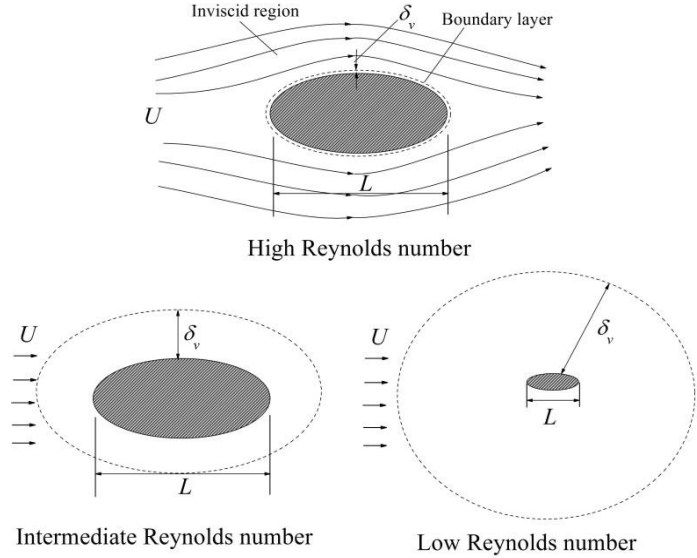

Figure 4 The schematic demonstration of physical meanings of Reynolds number

The $Re$ can be defined by the inertia force and viscous force.

$$Re = \frac{UL}{\nu} = \frac{L^2}{\nu L/U} = \frac{L^2}{\delta_v^2} = \frac{\text{inertia force}}{\text{viscous force}} \qquad (85)$$

where $\delta_v$ is the viscous diffusion length in the advection time interval of $L/U$.

$$\delta_v = \sqrt{\nu L/U} \qquad (86)$$

$Re$ is one of the most important parameter governing fluid dynamics. With different value of $Re$, fluid flow shows drastically different characteristics.

For simplicity, we consider the steady flow case where the gravitational force has no consequence to the dynamic of the flow. Then, the pressure $P$ represents the dynamics pressure. As shown in Figure 4, there are three kinds of flows, high Reynolds number flow ($Re \gg 1$), intermediate number flow ($Re \approx 1$), and low Reynolds number flow ($Re \ll 1$). Viscous force is unimportant outside region of O($\delta_v$) (inviscid region), but can become very important inside the region O($\delta_v$) close to the solid boundary (viscous region).

When $Re \gg 1$ (e.g. $\gg 10^3$ for common flow), inertia force is much greater than viscous force, i.e. the viscous diffusion distance is much less than the length $L$. This means that the viscous region is confined to a layer relatively thinner than the length scale $L$. Most of the region in the flow domain is inviscid as shown in Figure 4. This region near the solid boundary, where viscous term is important, is called a boundary layer. At high $Re$, flow instability may occur and turbulence is likely to appear in this high $Re$ flow regime. Flow is inviscid in the region outside the boundary layer where viscous force is negligible.

For high $Re$ flow, viscous terms in the normalized dimensionless equation can be dropped due to smallness. The first order approximation for ion species $i$ is:

$$0 = -\frac{P}{\rho_m U^2}\nabla^* p_i^* - \frac{1}{X_i}C_i^*\nabla^* \Phi_2^* \qquad (87)$$

The governing equation of momentum conservation in the inviscid flow region becomes

$$0 = -\nabla p_i - z_i F C_i \nabla \Phi_2 \qquad (88)$$

Flows in the boundary layer are still affected by viscous force, and new equations containing viscous term are required to account for the flow in the boundary layer

When $Re$ is close to 1, inertia forces and viscous forces are of equal importance. The flow is viscous in a region of $\delta_v$ surrounding the body. The height of is $\delta_v$ close to $L$ as shown in Figure 4. In this case, the viscous region is of similar size compared with the interested length scale $L$. No simplification can be done.

When $Re \ll 1$, the inertia force is very much smaller than the viscous force. The viscous diffusion length $\delta_v$ is much larger than $L$. The flow is viscous for almost entire region in the system except at positions very far away from the solid boundary. Figure 4 shows the flows with low Reynolds number.

When $Re \ll 1$, it implies the inertia force is negligible and the pressure force has to balance the viscous force. Therefore, the proper scale for $P$ is such that

$$\frac{P}{\rho_m U^2} = \frac{\mu}{\rho_m U L} \rightarrow P = \frac{\mu U}{L} \qquad (89)$$

Flows of low Reynolds number are called the Stokes flows, or creeping flows. One such example is the settling of small particles. The relatively small size of particles (compared with $\delta_v$) causes the Reynolds number to become very small.

6.3 Peclet number

The diffusion of the thermal energy from a hot solid wall is very similar to the diffusion of momentum in the boundary layer. Analogous to the hydrodynamic viscous diffusion length $\delta_v$, a thermal conduction length $\delta_T$ is defined as the region where heat conduction is important. The Peclet number $Pe$ is equal to

$$Pe = \frac{U_\infty L}{\alpha} = \frac{L^2}{\alpha L/U_\infty} = \left(\frac{L}{\delta_T}\right)^2 = \frac{convective\ heat\ transfer}{conductive\ heat\ transfer} \qquad (90)$$

where $\delta_T$ represents the thermal diffusion length (distance) during the time period when the fluids at velocity $U$ flowing from the leading edge (front tip) to the trailing edge (end) of the plate. The domain of distance O($\delta_T$) from the flat plate represents the thermal diffusion region. Outside the thermal diffusion region is the region dominated by pure convection.

We now examine the case when $Pe \ll 1$, i.e., $\delta_T \gg L$ when the thermal diffusion region has a dimension much larger than $L$. In the region of the distance O($L$) from the heated plate, the heat transfer is dominated by conduction. Consequently, the convection terms can be dropped from the energy equation, which to the first order approximation becomes:

$$\nabla^{*2} T^* = 0 \qquad (91)$$

Then, the heat convective effect represents only the higher order correction to the solution from the pure conduction.

As the $Pe$ increases from $Pe \ll 1$ to $Pe =$ O(1), the thermal diffusion length shrinks from $\delta_T \gg L$ to $\delta_T \approx$ L. Under this condition, both convection and conduction are of equally important in the region of O($L$) around the flat plate. Hence all terms have to be retained in the energy equations. No analytic solutions for the temperature are possible, except numerical solutions.

When the Peclet number is increased further from $Pe =$ O(1) to $Pe \gg 1$. The thermal diffusion length shrinks further from $\delta_T \approx$ L to $\delta_T \ll$ L. When $Pe \gg 1$, the conduction region becomes a thin layer wrapping around the surface boundary of the heated solid wall. This thin layer is called the thermal boundary layer. In the region of the geometric length scale of L, the conduction is not important and the heat transfer is due to pure convection. However, since the incoming fluid is at a temperature of $T_2$, the temperature in the pure convection region is $T_2$ everywhere which is of little interest. Apparently, all the important phenomena occur in the thermal boundary layer.

6.4 Prandtl number

The Prandtl number (*Pr*) is defined by following equation:

$$Pr = \frac{\nu}{\alpha} = (\frac{\delta_v}{\delta_T})^2 = \frac{momentum\ diffusion}{thermal\ diffusion} \qquad (92)$$

Prandtl number reflects the ratio of the lengths between the viscous layer and thermal diffusion layer. Therefore, we are more interested in two limit case. The case of $Pr \ll 1$ corresponds to $\delta_T \gg \delta_v$, where thermal diffusion is much faster. For the case of $Pr \gg 1$ corresponds to $\delta_v \gg \delta_T$.

6.5 Peclet number for mass transfer

The Peclet number for mass ($Pe_m$) is defined as

$$Pe_m = \frac{\rho_m U L}{\lambda_i} = \frac{convective\ mass\ transfer}{conductive\ mass\ transfer} \qquad (93)$$

The Peclet number for mass ($Pe_m$) represents the ratio of mass transfer between the convection and conduction. With a very high $Pe_m$ number the mass is transferred mainly by convection, while with a very low $Pe_m$ number the mass is transferred mainly by conduction. When $Pe_m$ number is medial, the convection and conduction are equally important.

6.6 *X* number

The *X* number is a new number. For species *i* it is defined as

$$X_i = \frac{\rho_m U^2/L}{z_i F C_i^0 \Phi_\Theta/L} = \frac{inertia\ force}{electric\ force} \qquad (94)$$

This equation indicates how much the ion species is influenced by the electric force when the ion species with the velocity of *U* passes through a headed solid landscape of *L*. With a very high *X* number, the ion species will be slightly influenced by the electric forces and it will pass through the solid landscape driven by the inertia force. While with a very low *X* number the ion species will be attracted (or repelled) by the electric force. Let's take the attractive electric force as an example. With a very low $X$ number the ion species will be attracted by the electric force and then arrive at the interface between the solid and the electrolyte to participate in the electrochemical reaction or formation of electrochemical double layer. At a medial $X$ number, the inertia force and electric force are equally important.

## 7 Concluding remarks

We established the general theory to describe the electrochemical flows. We derived the governing equations from the conservation laws of charge, mass, momentum, energy and mass concentration in the microscopic REV. In addition, we normalize the governing equations to derive the dimensionless parameters, known as Reynolds, Thompson, Peclet, Prandtl and $X$ numbers. The physical meaning of these parameter numbers in the electrochemical flow is discussed in detail. For example, the Thompson number represents the relative magnitude between two time scales, the advective time ($\tau_a$) and the response time ($\tau$). The electrochemical flow can be divided into transient or quasi-steady flow based on the magnitude of Thompson number. A new number, named $X$ number, appears in the Navier-Stokes equation symbolizing the balance between the inertia force and the electric force. Our general theory provides an approach to describe the motion of the electrochemical flows.

**Acknowledgements**

The author C. Xu would like to thank Dr. Sum Wai Chiang for helpful discussion. We like to thank the financial support from National Nature Science Foundation of China under Grants (No. 51102139) and from Shenzhen Technical Plan Projects (No. JC201105201100A). We also thank the financial support from Guangdong Province Innovation R&D Team Plan (2009010025). CT would like to thank City Key Laboratory of Thermal Management Engineering and Materials for financial support.

**Appendix**

| | |
|---|---|
| Notation | **v** velocity |
| REV representative elementary volume | $\nabla$ Hamilton operator |
| $f$ physical properties | $\nabla^2$ Laplacian operator |
| **r** location | $i$ one species |
| $t$ time | $dx_i$ scale length of microscopic REV |

$x_1$ one axis of the Cartesian coordinate

$x_2$ one axis of the Cartesian coordinate

$x_3$ one axis of the Cartesian coordinate

$V$ volume

$V_c$ volume of the REV

$\rho_m$ mass density of electrolyte

$\rho_{mi}$ mass density of species $i$

$N$ total number of the molecules and ions

$N_i$ number of species $i$

$m_i$ mass of one particle of species $i$

$\rho_c$ total charge density

$\rho_{ci}$ charge density of species $i$

$z_i$ elementary charge of species $i$

$n$ total number density

$n_i$ number density of species $i$

$C_i$ concentration of species $i$

$C_i^0$ concentration of species $i$ in the bulk electrolyte

$x_i$ concentration fraction of species $i$

$M_i$ atomic weight of species $i$

$F$ Faradic constant

$\mathbf{M}$ total translation momentum per unit volume of electrolyte

$\mathbf{v}_i$ velocity of one particle of species $i$

$u_i$ component of velocity of one particle of species $i$ at $x_1$

$v_i$ component of velocity of one particle of species $i$ at $x_2$

$w_i$ component of velocity of one particle of species $i$ at $x_3$

$\mathbf{u}$ mean velocity of the electrolyte fluid

$\mathbf{M}_i$ translation momentum of species $i$

$\mathbf{u}_s$ mean velocity of the solvent group

$\mathbf{u}_i$ mean velocity of species $i$

$\mathbf{v}_i'$ thermal velocity of species $i$

$T$ temperatures

$T_i$ temperatures of species $i$

$H_r$ heat per unit volume generated inside the REV

$\mathbf{E}_{ri}$ electric flied generated by ion species $i$

$\Phi_{ri}$ potential of ion species $i$

$\mathbf{E}_r$ electric potential generated inside the REV

$\mathbf{E}$ total electric flied through the REV surface

$\Phi_r$ potential of the REV

$Q_i$ total charge of species $i$ in the REV

$\mathbf{J}_i$ current flux of species $i$

$\mathbf{J}_0$ exchange current flux

$\mathbf{J}$ current flux

$S$ surface area

$s$ unit surface

$\mathbf{I}$ unit surface tensor

$D_i$ mass diffusion coefficient of species $i$

$\dot{m}_{ci}$ mass source of species $i$

$\mathbf{S}$ strain rate tensor

$\mathbf{b}$ body force

$C_p$ thermal capacity of electrolyte fluid

$k$ thermal conductivity of electrolyte fluid

$\dot{N}_i$ mass diffusion flux of species $i$

$p$ pressure

$O(1)$ order one

$U$ original velocity of the electrochemical fluid

$T_2$ original temperature of the electrochemical fluid

$T_1$ temperature of solid phase

$P$ pressure on the surface of the solid phase toward the electrochemical fluid

$Th$ Thompson number

$Re$ Reynolds number

$X$ new number appearing in the Navior-Stokes equation

$Pe$ Peclet number

$Pe_m$ Peclet number for the mass transfer

$Pr$ Prandtl number

Greek letters

k Boltzmann constant

$\sigma$ stress tensor

$\sigma_i$ stress tensor of species $i$

$\sigma^\ominus$ standard stress tensor at thermal equilibrium

$\sigma_{ri}$ resistance stress tensor for the species $i$

$\sigma_p$ pressure stress tensor

$\tau$ vicious stress tensor

$\varepsilon$ dielectric constant of the electrolyte

$\lambda_i$ diffusion conductivity of species $i$

$\tau$ time scale of one electrochemical fluid lasting

$\tau_a$ advective time

$\Phi_1$ potential of solid phase

$\Phi_2$ potential of electrochemical fluid

$\Phi_d$ potential of the electrochemical double layer

$\eta$ overpotential of electrochemical reaction

$\sigma_d$ surface charge density of the electrochemical double layer

$\Phi_\Theta$ reference potential

ç volume viscosity

$\mu$ dynamic viscosity of the electrolyte

$v$ kinematic viscosity of the electrolyte

$\delta_v$ viscous diffusion length

$\delta_T$ thermal diffusion length

$\alpha$ thermal diffusivity

† Van't Hoff factor

Subscripts

1 solid phase

2 fluid phase

$i$ species

Superscripts

* new variable in the normalized governing equations